\documentclass[aps,prl,10pt,twocolumn,superscriptaddress,floatfix]{revtex4-1}
\usepackage{graphicx}
\usepackage{dcolumn}
\usepackage{bm}
\usepackage{amsmath}

\begin{document}

\title{MD simulation of the radiation influence on the thermophysical properties and the structure of water}

\author{L. A. Bulavin}
 \affiliation{Physics Faculty, Taras Shevchenko National University of Kyiv, 4 Glushkova Av.,  Kyiv, 03022, Ukraine}
\author{K. V. Cherevko}
 \affiliation{Physics Faculty, Taras Shevchenko National University of Kyiv, 4 Glushkova Av.,  Kyiv, 03022, Ukraine}
\author{D. A. Gavryushenko}
 \affiliation{Physics Faculty, Taras Shevchenko National University of Kyiv, 4 Glushkova Av.,  Kyiv, 03022, Ukraine}
\author{V. M. Sysoev}
 \affiliation{Physics Faculty, Taras Shevchenko National University of Kyiv, 4 Glushkova Av., Kyiv, 03022, Ukraine}
\author{T. S. Vlasenko}
 \email{vlasenko.tata@gmail.com}
 \affiliation{Institute for Safety Problems of Nuclear Power Plants, National Academy of Sciences of Ukraine, 12 Lysogirska St., Kyiv, 03028, Ukraine}

\date{\today}

\begin{abstract}
The results of the molecular dynamics simulation of the radiation influence on the structure and thermophysical properties of water are presented. The changes in the radial distribution functions, momentum distribution function and the selfdiffusion coefficients are quantified. It is shown that the irradiation causes the structural and thermodynamics properties of water. The "effective temperature" of the stationary nonequilibrium water system under the irradiation allowing to define the correspondent equilibrium system with the same structural and thermodynamic properties is calculated. It is confirmed that the structural changes in the liquid systems under irradiation are caused by the changes in the coefficients of the Maxwell distribution function due to the momentum exchange between the active particles and the particles forming the liquid. To explain the phenomena observed in the molecular dynamics simulation the results are quantitatively compared to the predictions of the theoretical model of the phenomenon that is based on the Bogolyubov chain of equations and with the experimental data.
\end{abstract}\texttt{}

\pacs{61.20.Ja 05.20.-y 05.70.-a 61.20.-p 61.80.Az}

\maketitle

Recently, quite a number of works devoted to the radiation interaction with the matter appeared in the literature \cite{Zarkadoula2013, Yuan2009a, Luna2005, Weber1998, Fielden1991, Phillips1958}. That field is important both from fundamental point of view and for the different applications. The studies may reveal the underlaying physical mechanisms of the interactions and their results may be used for developing new technologies in medicine, nuclear energy, etc. Nowadays most of the studies in the field deal with the solid state \cite{Trachenko2005, Trachenko2002, Nordlund1998}, ionic liquids \cite{Yuan2009, Qi2008} or biological structures \cite{Alizadeh2013, Spotheim-Maurizot2011}. For the above objects computer simulation studies has shown to be on of the most effective instruments \cite{Christie2015, Lumpkin2008, Trachenko2006} as the experiments with the irradiation influence on matter are difficult to perform and to interpret. At the same time there are quite few works devoted to the irradiation influence on liquid systems. It should be noted that computer simulations are widely used to study the equilibrium and nonequilibrium liquid systems \cite{Xu2012, Elfimova2013, Toxvaerd1998, Spoel1998, Robinson1996, Allen1987}. Therefore, it looks surprising that computer simulations are not used to study the radiation influence on liquids. Furthermore, at present, most of the works in the field deal with the physical-chemical stage and describe the radiolysis \cite{Draganic2005, Burns1989, Sims2006} but only few can be found that search for the physical mechanisms involved in the process. One should also mention the existing experimental studies of the radiation influence on liquid matter \cite{Kolesnichenko1975, Martino2006, Zenkievicz2007, Byung2008, Weon2008}. Those experiments show the strong dependence of the thermodynamic properties of liquids on the radiation. Therefore, accounting for the importance of understanding the behavior of liquids under irradiation for the different applications it seems to be attractive to study the physical nature of the interaction.

The aim of this work is to use the molecular dynamics (MD) methods to study the physical mechanisms responsible for the changes in the structural and thermodynamic properties of water under irradiation by $\alpha$ particles (He).

Earlier, based on the fundamental Bogolyubov chain of equations \cite{Bogolyubov1962} there was suggested the model relating the structural and thermophysical properties of the nonequilibrium liquid systems under irradiation in the stationary state \cite{Bulavin2016}.
Within that model it was suggested that the irradiation changed the structure of the liquid system. That obtained new structure of the nonequilibrium liquid system in the stationary state was characterized by a new parameter that is the {\textquotedblleft}effective temperature{\textquotedblright} $T_{eff}$. That allowed writing down the classical BGY equation \cite{Temperley1968} for the nonequilibrium system in the stationary state:
\begin{multline}
\label{Ef1}
kT_{eff}\frac{\partial F_{2}({{\bf r}_1},{{\bf r}_2})}{\partial {\bf r}_1}+\frac{\partial \Phi(|{{\bf r}_1}-{{\bf r}_2}|)}{\partial {\bf r}_1}F_{2}({{\bf r}_1},{{\bf r}_2})\\
+\rho\int\frac{\partial \Phi(|{{\bf r}_1}-{{\bf r}_3}|)}{\partial {\bf r}_1}F_{3}({{\bf r}_1},{{\bf r}_2},{{\bf r}_3})d{\bf r}_3=0,
\end{multline}
with $T_{eff}$ standing instead of $T$. Here $F_2({{\bf r}_1},{{\bf r}_2})$ and $F_3({{\bf r}_1},{{\bf r}_2},{{\bf r}_3})$ are the 2-nd and 3-d order distribution functions respectively depending on space coordinates ${{\bf r}_1},{{\bf r}_2},{{\bf r}_3}$, $\Phi(|{{\bf r}_i}-{{\bf r}_{j}}|)$ \t is the potential of interaction between $i$-th and $j$-th particles, $\rho=\frac {N} {V}$ \t is the numerical density.
The important thing to mention is the fact that the introduced parameter $T_{eff}$ was, actually, the analogue of the thermodynamic temperature for the nonequilibrium system in focus. It was suggested to be equal to the thermodynamic temperature of the corresponding equilibrium system with the same structural properties. It was shown that the {\textquotedblleft}effective temperature{\textquotedblright} could be calculated from the perturbed momentum distribution functions of the systems caused by the momentum exchange between the active particles and the particles forming the liquid.
\begin{equation}
\label{EfT}
kT_{eff}\int\frac{\partial {f_2}({{\bf p}_1},{{\bf p}_2})}{\partial{\bf p}_1}d{\bf p}_1d{\bf p}_2=-\int\frac{{\bf p}_1}{m}{f_2}({{\bf p}_1},{{\bf p}_2})d{{\bf p}_1}d{{\bf p}_2}.
\end{equation}
Therefore, the developed theoretical model of the process \cite{Bulavin2016} allowed calculating thermophysical properties of the stationary nonequilibrium liquid systems under irradiation from the thermophysical properties of the correspondent equilibrium liquid systems considering the new parameter $T_{eff}$.

In this work we present the results of the MD simulation of the radiation influence on water and compare them with the predictions of the theoretical model suggested earlier.

The MD simulations using the rigid simple point charge (SPC) model of water \cite{Berendsen1981} are performed for a cubic simulation box containing 16384 water molecules and 1 He particle at the temperature 300 K, and at ambient pressure (0.1 MPa). The geometric parameters of the SPC model used are given in Tab \ref{spcparam}.
\begin{table}[htp]
\caption{Geometric parameters of the rigid SPC model used \cite{Wu2006}}
\label{spcparam}
 \begin{ruledtabular}
 \begin{tabular}{lc}
 $\theta^o_{\angle{HOH}}$ & 109.47\\
 $r_{OH}$, ${\AA}$ & 1 \\
 $r_{HH}$, ${\AA}$ & 1.63 \\
 \end{tabular}
\end{ruledtabular}
\end{table}
The DL POLY package (4.06 version) is used \cite{Todorov2006}, with the Ewald particle mesh method for the evaluation of the Coulomb interactions. The intermolecular potential energy between atomic sites is calculated in a standard way by a sum of the Lennard-Jones (12-6) potential and the Coulomb electrostatic interaction
\begin{multline}
\label{Potential}
U\left( {{r}_{ij}} \right)={{U}_{LJ}}\left( {{r}_{ij}} \right)+{{U}_{Coul}}\left( {{r}_{ij}} \right)\\
=\sum\limits_{i\prec j}{4{{\varepsilon }_{ij}}\left( {{\left( \frac{{{\sigma }_{ij}}}{{{r}_{ij}}} \right)}^{12}}-{{\left( \frac{{{\sigma }_{ij}}}{{{r}_{ij}}} \right)}^{6}} \right)}+\sum\limits_{i\prec j}{\frac{{{q}_{i}}{{q}_{j}}}{4\pi {{\varepsilon }_{0}}{{r}_{ij}}}}
\end{multline}
with the off-diagonal interaction parameters calculated using the Lorentz-Berthelot mixing rules \cite{Lorentz1881, Berthelot1898}
\begin{align}
\label{Parameters}
{{\sigma }_{ij}}=\frac{{{\sigma }_{i}}+{{\sigma }_{j}}}{2},\\
{{\varepsilon }_{ij}}=\sqrt{{{\varepsilon }_{i}}{{\varepsilon }_{j}}}.
\end{align}

The parameters used are given in Tab. \ref{spcparam1}.
\begin{table}[htp]
\caption{Masses and intermolecular potential parameters of water and He \cite{Zielkiewicz2005, Lyubartsev1996, Tang2006}}
\label{spcparam1}
 \begin{ruledtabular}
 \begin{tabular}{lcccc}
 Atom & $m$, a.u.m. & $q$, $e$ & $\varepsilon$, $kJ\cdot{mol}^{-1}$ & $\sigma$, ${\AA}$\\
 H & 1 & +0.41 & 0 & 0\\
 O & 16 & -0.82 & 0.65 & 3.1656\\
 He & 4.0026 & +2 & 0.084 & 2.6\\
 \end{tabular}
\end{ruledtabular}
\end{table}
All the simulations are performed in two steps. First, the system is equilibrated in the {\textit{NVT}} ensemble using Berendsen thermostat and the box is fixed. In all the future simulation the box size remains unchanged. Next, the radiation is switched on. At this stage the simulation is performed in the {\textit{NVE}} ensemble. Irradiation is included by accelerating the He atom originally present in the system. In order to reach the stationary state of the system we add energy in a discrete way with the step 0.05 keV. It is done by accelerating He every 2 $ps$ to have the total irradiation energy ranging from 0.05 keV to 0.25 keV.

To justify the choice of the {\textit{NVE}} ensemble one can mention the fact that in all the other ensembles inclusion of the thermostats makes the momentum distribution function move towards the Maxwellian one and, hence, disturbs the real physical picture in the momentum space. Therefore, they ar not acceptable for our case as one expect the changes in the momentum distribution function to be one of the main physical mechanisms of the radiation influence on he structural and thermodynamic properties of water. All the parameters of the simulation process are given in Tab \ref{simparam}.

\begin{table}[htp]
\caption{Simulation parameters}
\label{simparam}
 \begin{ruledtabular}
 \begin{tabular}{lc}
 Number of water molecules & 16384 \\
 Box size,${\AA}$  & 79 \\
 Density $\rho$, $kg\cdot{m}^{-3}$ \cite{Lemmon2007} & 992,27 \\
 Temperature {\textit{T}}, K & 300 \\
 Equilibration time ({\textit{NVT}} ensemble), ps  & 2000 \\
 Simulation time ({\textit{NVE}} ensemble), ps  & 2-10 \\
 $R_{cutoff}$, $\AA$  & 14 \\
 Boundary conditions  & periodic \\
 Simulation software  & DL-POLY 4.06 \\
 \end{tabular}
\end{ruledtabular}
\end{table}

The SPC rigid model was used in our simulations as it gives structural and dynamic parameters of the bulk water that are in accord with the experimental data \cite{Zielkiewicz2005, Guillot2002} and at  the same time it gives the results for the radial distribution functions (RDF) and momentum distribution functions that are not blurred by inclusion of the internal degrees of freedom. It seems to be attractive as we suggest that one of the main mechanisms responsible for the changes of the thermodynamic properties of the liquid systems under irradiation is the change in the momentum distribution function \cite{Bulavin2016}. Therefore, using SPC model seems to be a reasonable choice in  the first approximation.

\begin{figure}[h]
\center{\includegraphics[scale=1.0]{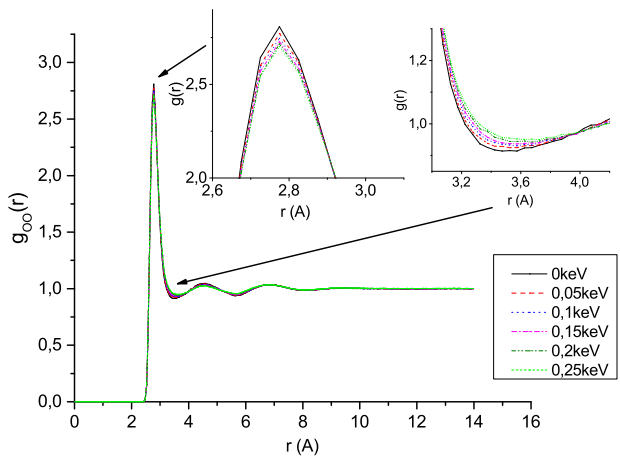}}
\caption{Oxygen-oxygen RDFs for the radiation energies ranging from 0 KeV to 0.25 KeV}
\label{OORDF}
\end{figure}

\begin{figure}[h]
\center{\includegraphics[scale=1.0]{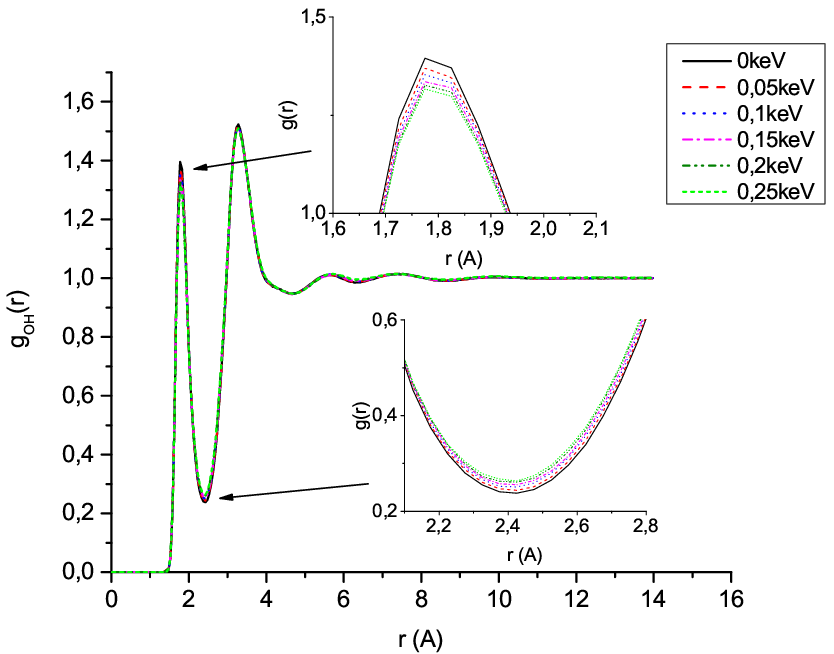}}
\caption{Oxygen-hydrogen RDFs for the radiation energies ranging from 0 KeV to 0.25 KeV}
\label{OHRDF}
\end{figure}

\begin{figure}[h]
\center{\includegraphics[scale=1.0]{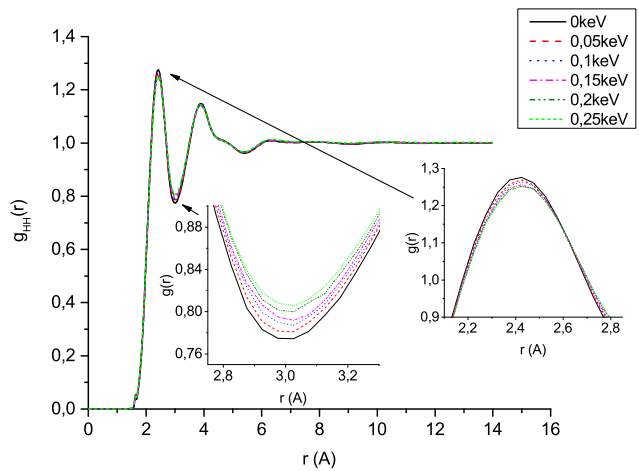}}
\caption{Hydrogen-hydrogen RDFs for the radiation energies ranging from 0 KeV to 0.25 KeV}
\label{HHRDF}
\end{figure}

In order to have the reliable data for the noneqilibrium system in the stationary state we have averaged the results of the three runs with the same initial equilibrated system. The $g_{OO}$, $g_{OH}$, and $g_{HH}$ RDFs are shown at Figs. \ref{OORDF}, \ref{OHRDF}, and \ref{HHRDF} respectively.

It can be seen from the figures that all the changes in the RDFs are quite small. It can be explained by the averaging over the three different runs. Still, it can be easily seen that there is the correlation in between the radiation energy and the changes in the RDFs.
At the inset of Fig. \ref{OORDF} one can see the decrease of the height of the first peak with the increasing radiation energy. Such a behavior suggests the tendency to blurring of the first and second coordination spheres.  At the same time its position remains the same at about 2.78 $\AA$ that corresponds to the geometric criteria for the hydrogen bond that is $R_{OO}\leq3.3 {\AA}$ \cite{Robinson1996}.

For $g_{OH}(r)$ the decrease of the height of the first peak with the increasing radiation energy(the inset at Fig.\ref{OHRDF}) is seen again. Its position at 1.8 $\AA$ corresponds to the geometric criteria for the hydrogen bond that is $R_{OH}\leq2.6 {\AA}$ \cite{Robinson1996}. Such a behavior suggests the tendency of the radiation to destroy the net of hydrogen bonds.

Unfortunately, the changes in the $g_{HH}(r)$ are hardly seen (Fig.\ref{HHRDF}). Still, the tendency of the first peak to decrease with the increasing radiation energy is also observed at the graph.

\begin{figure}[h]
\center{\includegraphics[scale=1.0]{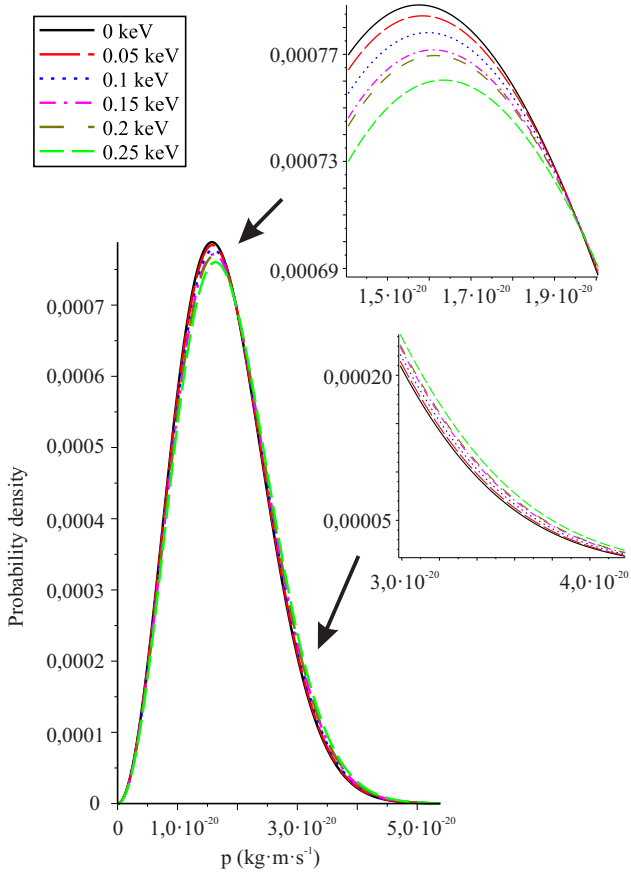}}
\caption{Momentum distribution functions for the radiation energies ranging from 0 KeV to 0.25 KeV}
\label{Momdistr}
\end{figure}

To find the physical mechanism of the radiation influence on water causing the observed structural changes the obtained simulation results are compared to the theoretical model of the process developed earlier \cite{Bulavin2016}. In order to calculate the {\textquotedblleft}effective temperatures{\textquotedblright} that characterize the structural and thermophysical properties of the stationary nonequilibrium liquid system under irradiation the momentum distribution functions have been extracted from the simulation data for the different irradiation energies. The obtained curves approximated with the Maxwell type functions $f(p)=A{\exp(-\phi{p}^2)}$ are shown at Fig. \ref{Momdistr}. One can see that the position of the maximum shifts toward the higher velocities and the peak broadens with the increasing radiation energy. Such a behavior qualitatively confirms our hypothesis that one of the important mechanisms of the irradiation influence on the properties of water is the change of the velocity distribution function due to the momentum exchange between the active particles and the particles forming the liquid.

In order to find the thermophysical properties of the system we calculate the {\textquotedblleft}effective temperature{\textquotedblright} $T_{eff}$ from the equation (\ref{EfT}) with the correspondent momentum distribution functions ${f}(\mid{\bf p}\mid)$ shown at Fig. \ref{Momdistr}. The resulting values are given in Tab. \ref{EfTval}
\begin{table}[htp]
\caption{Effective temperatures for the different irradiation energies}
\label{EfTval}
 \begin{ruledtabular}
 \begin{tabular}{lcccccc}
 Irradiation energy, keV & 0 & 0.05 & 0.1 & 0.15 & 0.2 & 0.25 \\
 $T_{eff}$, K & 300.2 & 303.4 & 308.3 & 313.5 & 315.2 & 322.8\\
 $T_{irrad}$, K & 300 & 312 & 324 & 335 & 347 & 359\\
 \end{tabular}
\end{ruledtabular}
\end{table}
As our system is in the stationary nonequilibrium state it is not possible to define the {\textquotedblleft}real{\textquotedblright} temperature of the system. Therefore, to have some data for comparison in the last raw of Tab. \ref{EfTval} there are the temperatures that should have the system if all the radiation energy goes for heating only. It can be seen that the numbers are different. At this point it is important to mention that the RDFs of the equilibrium system calculated in the NVT ensemble at real temperatures equal to $T_{eff}$ given in Tab. \ref{EfTval} coincide with the RDFs shown at Figs. \ref{OORDF}-\ref{HHRDF} for the corresponding irradiation energies.

\begin{figure}[h]
\center{\includegraphics[scale=1.0]{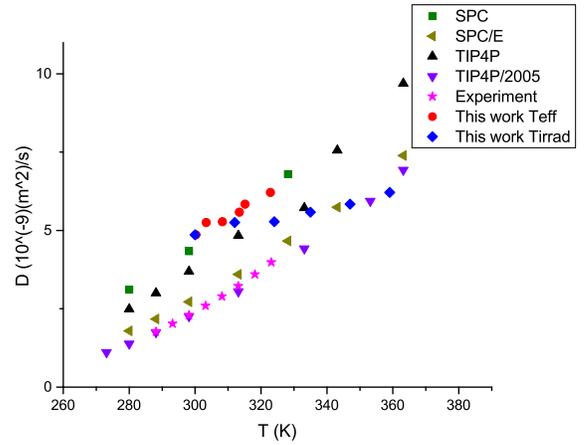}}
\caption{Selfdiffusion coefficient dependence on temperature: Data from literature for the different models and experiment \cite{GuevaraCarrion2011}; Dependence obtained in this work from $T_{eff}$ (circles) and from $T_{irrad}$ (diamonds).}
\label{SD}
\end{figure}

To check the hypothesis that it is the effective temperature but not the {\textquotedblleft}real{\textquotedblright} or $T_{irrad}$ that describes the thermophysical properties of the system under the irradiation it is intereesting to compare simulation results with the experimental data. The effects might be seen in all the thermophysical properties that are defined by the structure of the system like surface tension coefficient, selfdiffusion coefficient, {\textit{etc.}}. We have analyzed the selfdiffusion coefficient dependence on the effective temperature. The corresponding graph is shown at  Fig. (\ref{SD} ). At the same figure there are shown graphs of the selfdiffusion coefficient dependence on real temperature that can be found in the literature \cite{GuevaraCarrion2011}.
One can see from Fig. \ref{SD} that the obtained results for the selfdiffusion coefficient plotted against the $T_{eff}$ are in accord with the data available in the literature for the SPC rigid model. At the same time calculating the {\textquotedblleft}real{\textquotedblright} temperature for the system in the nonequilibrium state is not possible and plotting $D$ against $T_{irrad}$ (Fig. \ref{SD}) doesn't show correspondence to the existing data. Therefore, one can come to the conclusion that it is the {\textquotedblleft}effective{\textquotedblright} temperature that can explain the observed behavior of the selfdiffusion coefficient.

In this work the results of the MD simulations of the $\alpha$-particles radiation influence on water are presented. It is shown that irradiation causes the changes in the structural and thermodynamic properties of water. The main observed changes in the structure are the blurring of the first and the second coordination spheres as well as destruction of the net of hydrogen bonds.

The comparison of the obtained results with the predictions of the earlier developed theoretical model of the process based on the Bogolyubov chain of equations shows that the observed changes in structure are characterized by a new parameter that is the {\textquotedblleft}effective temperature{\textquotedblright}. It means that the structure of the liquid system under irradiation in the stationary nonequilibrium state is the same as the structure of the corresponding equilibrium system with its thermodynamic temperature equal to the {\textquotedblleft}effective temperature{\textquotedblright}. Therefore, the {\textquotedblleft}effective temperature{\textquotedblright} might be treated as the {\textquotedblleft}structural temperature{\textquotedblright} of the nonequilibrium system in the stationary state.

The changes of the selfdiffusion coefficient under irradiation are shown to be caused by the changes in the structure of the liquid system that are described by the growth of the {\textquotedblleft}effective temperature{\textquotedblright}. The dependence of the selfdiffusion coefficient on the {\textquotedblleft}effective temperature{\textquotedblright} shows the good correspondence to the existing data for the temperature dependence of the selfdiffusion coefficient. From the obtained results it is seen that selfdiffusion coefficient grows with the increasing radiation energy. At the same time the changes in the {\textquotedblleft}real temperature {\textquotedblright} of the nonequilibrium system can not explain the observed growth of the selfdiffusion coefficient in a correct way. Therefore, it is shown that the thermophysical properties of the liquid system under irradiation in the stationary nonequilibrium state are the same as the thermophysical properties of the corresponding equilibrium system with its thermodynamic temperature equal to the {\textquotedblleft}effective temperature{\textquotedblright}.

It is shown that the changes in the coefficients of the Maxwell distribution function due to the momentum exchange between the active particles and the particles forming the liquid is one of the important physical mechanisms of the radiation influence on liquid systems. The knowledge of the distorted momentum distribution function of the liquid system under irradiation in the stationary nonequilibrium state allows calculating the {\textquotedblleft}effective temperature{\textquotedblright} of such a system and, hence, its structural and thermodynamic properties.

The obtained in the work results of the MD simulation of the radiation influence on water quantitatively confirm the predictions of the previously introduced theoretical model of the process and are in accord with the available in the literature experimental data.

\bibliography{IrradLiquids}

\begin{thebibliography}{44}%
\makeatletter
\providecommand \@ifxundefined [1]{%
 \@ifx{#1\undefined}
}%
\providecommand \@ifnum [1]{%
 \ifnum #1\expandafter \@firstoftwo
 \else \expandafter \@secondoftwo
 \fi
}%
\providecommand \@ifx [1]{%
 \ifx #1\expandafter \@firstoftwo
 \else \expandafter \@secondoftwo
 \fi
}%
\providecommand \natexlab [1]{#1}%
\providecommand \enquote  [1]{``#1''}%
\providecommand \bibnamefont  [1]{#1}%
\providecommand \bibfnamefont [1]{#1}%
\providecommand \citenamefont [1]{#1}%
\providecommand \href@noop [0]{\@secondoftwo}%
\providecommand \href [0]{\begingroup \@sanitize@url \@href}%
\providecommand \@href[1]{\@@startlink{#1}\@@href}%
\providecommand \@@href[1]{\endgroup#1\@@endlink}%
\providecommand \@sanitize@url [0]{\catcode `\\12\catcode `\$12\catcode
  `\&12\catcode `\#12\catcode `\^12\catcode `\_12\catcode `\%12\relax}%
\providecommand \@@startlink[1]{}%
\providecommand \@@endlink[0]{}%
\providecommand \url  [0]{\begingroup\@sanitize@url \@url }%
\providecommand \@url [1]{\endgroup\@href {#1}{\urlprefix }}%
\providecommand \urlprefix  [0]{URL }%
\providecommand \Eprint [0]{\href }%
\providecommand \doibase [0]{http://dx.doi.org/}%
\providecommand \selectlanguage [0]{\@gobble}%
\providecommand \bibinfo  [0]{\@secondoftwo}%
\providecommand \bibfield  [0]{\@secondoftwo}%
\providecommand \translation [1]{[#1]}%
\providecommand \BibitemOpen [0]{}%
\providecommand \bibitemStop [0]{}%
\providecommand \bibitemNoStop [0]{.\EOS\space}%
\providecommand \EOS [0]{\spacefactor3000\relax}%
\providecommand \BibitemShut  [1]{\csname bibitem#1\endcsname}%
\let\auto@bib@innerbib\@empty
\bibitem [{\citenamefont {Zarkadoula}\ \emph {et~al.}(2013)\citenamefont
  {Zarkadoula}, \citenamefont {Daraszewicz}, \citenamefont {Duffy},
  \citenamefont {Seaton}, \citenamefont {Todorov}, \citenamefont {Nordlund},
  \citenamefont {Dove},\ and\ \citenamefont {Trachenko}}]{Zarkadoula2013}%
  \BibitemOpen
  \bibfield  {author} {\bibinfo {author} {\bibfnamefont {E.}~\bibnamefont
  {Zarkadoula}}, \bibinfo {author} {\bibfnamefont {S.~L.}\ \bibnamefont
  {Daraszewicz}}, \bibinfo {author} {\bibfnamefont {D.~M.}\ \bibnamefont
  {Duffy}}, \bibinfo {author} {\bibfnamefont {M.~A.}\ \bibnamefont {Seaton}},
  \bibinfo {author} {\bibfnamefont {I.~T.}\ \bibnamefont {Todorov}}, \bibinfo
  {author} {\bibfnamefont {K.}~\bibnamefont {Nordlund}}, \bibinfo {author}
  {\bibfnamefont {M.~T.}\ \bibnamefont {Dove}}, \ and\ \bibinfo {author}
  {\bibfnamefont {K.}~\bibnamefont {Trachenko}},\ }\href {\doibase
  10.1088/0953-8984/25/12/125402} {\bibfield  {journal} {\bibinfo  {journal}
  {J. Phys.: Condens. Matter}\ }\textbf {\bibinfo {volume} {25}},\ \bibinfo
  {pages} {125402} (\bibinfo {year} {2013})}\BibitemShut {NoStop}%
\bibitem [{\citenamefont {Yuan}\ \emph
  {et~al.}(2009{\natexlab{a}})\citenamefont {Yuan}, \citenamefont {Peng},
  \citenamefont {Xu}, \citenamefont {Zhai}, \citenamefont {Li},\ and\
  \citenamefont {Wei}}]{Yuan2009a}%
  \BibitemOpen
  \bibfield  {author} {\bibinfo {author} {\bibfnamefont {L.}~\bibnamefont
  {Yuan}}, \bibinfo {author} {\bibfnamefont {J.}~\bibnamefont {Peng}}, \bibinfo
  {author} {\bibfnamefont {L.}~\bibnamefont {Xu}}, \bibinfo {author}
  {\bibfnamefont {M.}~\bibnamefont {Zhai}}, \bibinfo {author} {\bibfnamefont
  {J.}~\bibnamefont {Li}}, \ and\ \bibinfo {author} {\bibfnamefont
  {G.}~\bibnamefont {Wei}},\ }\href {\doibase 10.1021/jp9016079} {\bibfield
  {journal} {\bibinfo  {journal} {J. Phys. Chem. B}\ }\textbf {\bibinfo
  {volume} {113}},\ \bibinfo {pages} {8948–8952} (\bibinfo {year}
  {2009}{\natexlab{a}})}\BibitemShut {NoStop}%
\bibitem [{\citenamefont {Luna}\ and\ \citenamefont
  {Montenegro}(2005)}]{Luna2005}%
  \BibitemOpen
  \bibfield  {author} {\bibinfo {author} {\bibfnamefont {H.}~\bibnamefont
  {Luna}}\ and\ \bibinfo {author} {\bibfnamefont {E.~C.}\ \bibnamefont
  {Montenegro}},\ }\href {\doibase
  https://doi.org/10.1103/PhysRevLett.94.043201} {\bibfield  {journal}
  {\bibinfo  {journal} {Phys. Rev. Lett.}\ }\textbf {\bibinfo {volume} {94}},\
  \bibinfo {pages} {043201} (\bibinfo {year} {2005})}\BibitemShut {NoStop}%
\bibitem [{\citenamefont {Weber}\ \emph {et~al.}(1998)\citenamefont {Weber},
  \citenamefont {Ewing}, \citenamefont {Catlow}, \citenamefont {de~la Rubia},
  \citenamefont {Hobbs}, \citenamefont {Kinoshita}, \citenamefont {Matzke},
  \citenamefont {Motta}, \citenamefont {Nastasi}, \citenamefont {Salje},
  \citenamefont {Vance},\ and\ \citenamefont {Zinkle}}]{Weber1998}%
  \BibitemOpen
  \bibfield  {author} {\bibinfo {author} {\bibfnamefont {W.~J.}\ \bibnamefont
  {Weber}}, \bibinfo {author} {\bibfnamefont {R.~C.}\ \bibnamefont {Ewing}},
  \bibinfo {author} {\bibfnamefont {C.~R.~A.}\ \bibnamefont {Catlow}}, \bibinfo
  {author} {\bibfnamefont {T.~D.}\ \bibnamefont {de~la Rubia}}, \bibinfo
  {author} {\bibfnamefont {L.~W.}\ \bibnamefont {Hobbs}}, \bibinfo {author}
  {\bibfnamefont {C.}~\bibnamefont {Kinoshita}}, \bibinfo {author}
  {\bibfnamefont {H.}~\bibnamefont {Matzke}}, \bibinfo {author} {\bibfnamefont
  {A.~T.}\ \bibnamefont {Motta}}, \bibinfo {author} {\bibfnamefont
  {M.}~\bibnamefont {Nastasi}}, \bibinfo {author} {\bibfnamefont {E.~K.~H.}\
  \bibnamefont {Salje}}, \bibinfo {author} {\bibfnamefont {E.~R.}\ \bibnamefont
  {Vance}}, \ and\ \bibinfo {author} {\bibfnamefont {S.~J.}\ \bibnamefont
  {Zinkle}},\ }\href {\doibase 10.1557/JMR.1998.0205} {\bibfield  {journal}
  {\bibinfo  {journal} {JMR}\ }\textbf {\bibinfo {volume} {13}},\ \bibinfo
  {pages} {1434} (\bibinfo {year} {1998})}\BibitemShut {NoStop}%
\bibitem [{\citenamefont {Fielden}\ and\ \citenamefont
  {O'Neill}(1991)}]{Fielden1991}%
  \BibitemOpen
  \bibinfo {editor} {\bibfnamefont {E.~M.}\ \bibnamefont {Fielden}}\ and\
  \bibinfo {editor} {\bibfnamefont {P.}~\bibnamefont {O'Neill}},\ eds.,\ \href
  {\doibase 10.1007/978-3-642-75148-6} {\emph {\bibinfo {title} {The Early
  Effects of Radiation on DNA}}},\ \bibinfo {series} {NATO ASI Series H: Cell
  Biology}, Vol.~\bibinfo {volume} {54}\ (\bibinfo  {publisher}
  {Springer-Verlag},\ \bibinfo {year} {1991})\BibitemShut {NoStop}%
\bibitem [{\citenamefont {Phillips}\ \emph {et~al.}(1958)\citenamefont
  {Phillips}, \citenamefont {Moody},\ and\ \citenamefont
  {Mattok}}]{Phillips1958}%
  \BibitemOpen
  \bibfield  {author} {\bibinfo {author} {\bibfnamefont {G.~O.}\ \bibnamefont
  {Phillips}}, \bibinfo {author} {\bibfnamefont {G.~J.}\ \bibnamefont {Moody}},
  \ and\ \bibinfo {author} {\bibfnamefont {G.~L.}\ \bibnamefont {Mattok}},\
  }\href {\doibase 10.1039/JR9580003522} {\bibfield  {journal} {\bibinfo
  {journal} {J. Chem. Soc.}\ ,\ \bibinfo {pages} {3522}} (\bibinfo {year}
  {1958})}\BibitemShut {NoStop}%
\bibitem [{\citenamefont {Trachenko}\ \emph {et~al.}(2005)\citenamefont
  {Trachenko}, \citenamefont {Pruneda}, \citenamefont {Artacho},\ and\
  \citenamefont {Dove}}]{Trachenko2005}%
  \BibitemOpen
  \bibfield  {author} {\bibinfo {author} {\bibfnamefont {K.}~\bibnamefont
  {Trachenko}}, \bibinfo {author} {\bibfnamefont {J.~M.}\ \bibnamefont
  {Pruneda}}, \bibinfo {author} {\bibfnamefont {E.}~\bibnamefont {Artacho}}, \
  and\ \bibinfo {author} {\bibfnamefont {M.~T.}\ \bibnamefont {Dove}},\ }\href
  {\doibase 10.1103/PhysRevB.71.184104} {\bibfield  {journal} {\bibinfo
  {journal} {Phys. Rev. B}\ }\textbf {\bibinfo {volume} {71}},\ \bibinfo
  {pages} {184104} (\bibinfo {year} {2005})}\BibitemShut {NoStop}%
\bibitem [{\citenamefont {Trachenko}\ \emph {et~al.}(2002)\citenamefont
  {Trachenko}, \citenamefont {Dove},\ and\ \citenamefont
  {Salje}}]{Trachenko2002}%
  \BibitemOpen
  \bibfield  {author} {\bibinfo {author} {\bibfnamefont {K.}~\bibnamefont
  {Trachenko}}, \bibinfo {author} {\bibfnamefont {M.~T.}\ \bibnamefont {Dove}},
  \ and\ \bibinfo {author} {\bibfnamefont {E.~K.~H.}\ \bibnamefont {Salje}},\
  }\href {\doibase 10.1103/PhysRevB.65.180102} {\bibfield  {journal} {\bibinfo
  {journal} {Phys. Rev. B}\ }\textbf {\bibinfo {volume} {65}},\ \bibinfo
  {pages} {180102(R)} (\bibinfo {year} {2002})}\BibitemShut {NoStop}%
\bibitem [{\citenamefont {Nordlund}\ \emph {et~al.}(1998)\citenamefont
  {Nordlund}, \citenamefont {Ghaly}, \citenamefont {Averback}, \citenamefont
  {Caturla}, \citenamefont {Diaz de~la Rubia},\ and\ \citenamefont
  {Tarus}}]{Nordlund1998}%
  \BibitemOpen
  \bibfield  {author} {\bibinfo {author} {\bibfnamefont {K.}~\bibnamefont
  {Nordlund}}, \bibinfo {author} {\bibfnamefont {M.}~\bibnamefont {Ghaly}},
  \bibinfo {author} {\bibfnamefont {R.~S.}\ \bibnamefont {Averback}}, \bibinfo
  {author} {\bibfnamefont {M.}~\bibnamefont {Caturla}}, \bibinfo {author}
  {\bibfnamefont {T.}~\bibnamefont {Diaz de~la Rubia}}, \ and\ \bibinfo
  {author} {\bibfnamefont {J.}~\bibnamefont {Tarus}},\ }\href {\doibase
  10.1103/PhysRevB.57.7556} {\bibfield  {journal} {\bibinfo  {journal} {Phys.
  Rev. B}\ }\textbf {\bibinfo {volume} {57}},\ \bibinfo {pages} {7556}
  (\bibinfo {year} {1998})}\BibitemShut {NoStop}%
\bibitem [{\citenamefont {Yuan}\ \emph
  {et~al.}(2009{\natexlab{b}})\citenamefont {Yuan}, \citenamefont {Peng},
  \citenamefont {Zhai}, \citenamefont {Li},\ and\ \citenamefont
  {Wei}}]{Yuan2009}%
  \BibitemOpen
  \bibfield  {author} {\bibinfo {author} {\bibfnamefont {L.}~\bibnamefont
  {Yuan}}, \bibinfo {author} {\bibfnamefont {J.}~\bibnamefont {Peng}}, \bibinfo
  {author} {\bibfnamefont {M.}~\bibnamefont {Zhai}}, \bibinfo {author}
  {\bibfnamefont {J.}~\bibnamefont {Li}}, \ and\ \bibinfo {author}
  {\bibfnamefont {G.}~\bibnamefont {Wei}},\ }\href {\doibase
  10.1016/j.radphyschem.2009.03.064} {\bibfield  {journal} {\bibinfo  {journal}
  {Rad. Phys. Chem.}\ }\textbf {\bibinfo {volume} {78}},\ \bibinfo {pages}
  {737–739} (\bibinfo {year} {2009}{\natexlab{b}})}\BibitemShut {NoStop}%
\bibitem [{\citenamefont {Qi}\ \emph {et~al.}(2008)\citenamefont {Qi},
  \citenamefont {Wua}, \citenamefont {Li},\ and\ \citenamefont {Luo}}]{Qi2008}%
  \BibitemOpen
  \bibfield  {author} {\bibinfo {author} {\bibfnamefont {M.}~\bibnamefont
  {Qi}}, \bibinfo {author} {\bibfnamefont {G.}~\bibnamefont {Wua}}, \bibinfo
  {author} {\bibfnamefont {Q.}~\bibnamefont {Li}}, \ and\ \bibinfo {author}
  {\bibfnamefont {Y.}~\bibnamefont {Luo}},\ }\href {\doibase
  10.1016/j.radphyschem.2007.12.007} {\bibfield  {journal} {\bibinfo  {journal}
  {Rad. Phys. Chem.}\ }\textbf {\bibinfo {volume} {77}},\ \bibinfo {pages}
  {877–883} (\bibinfo {year} {2008})}\BibitemShut {NoStop}%
\bibitem [{\citenamefont {Alizadeh}\ \emph {et~al.}(2013)\citenamefont
  {Alizadeh}, \citenamefont {Sanz}, \citenamefont {Garcia},\ and\ \citenamefont
  {Sanche}}]{Alizadeh2013}%
  \BibitemOpen
  \bibfield  {author} {\bibinfo {author} {\bibfnamefont {E.}~\bibnamefont
  {Alizadeh}}, \bibinfo {author} {\bibfnamefont {A.~G.}\ \bibnamefont {Sanz}},
  \bibinfo {author} {\bibfnamefont {G.}~\bibnamefont {Garcia}}, \ and\ \bibinfo
  {author} {\bibfnamefont {L.}~\bibnamefont {Sanche}},\ }\href {\doibase
  10.1021/jz4000998} {\bibfield  {journal} {\bibinfo  {journal} {Phys. Chem.
  Lett.}\ }\textbf {\bibinfo {volume} {4 (5)}},\ \bibinfo {pages} {820}
  (\bibinfo {year} {2013})}\BibitemShut {NoStop}%
\bibitem [{\citenamefont {Spotheim-Maurizot}\ and\ \citenamefont
  {Davidkova}(2011)}]{Spotheim-Maurizot2011}%
  \BibitemOpen
  \bibfield  {author} {\bibinfo {author} {\bibfnamefont {M.}~\bibnamefont
  {Spotheim-Maurizot}}\ and\ \bibinfo {author} {\bibfnamefont {M.}~\bibnamefont
  {Davidkova}},\ }\href {\doibase 10.1088/1742-6596/261/1/012010} {\bibfield
  {journal} {\bibinfo  {journal} {J. Phys.: Conf. Ser.}\ }\textbf {\bibinfo
  {volume} {261}},\ \bibinfo {pages} {012010} (\bibinfo {year}
  {2011})}\BibitemShut {NoStop}%
\bibitem [{\citenamefont {Christie}\ \emph {et~al.}(2015)\citenamefont
  {Christie}, \citenamefont {Robinson}, \citenamefont {Roach}, \citenamefont
  {Ross}, \citenamefont {Suarez-Martinez},\ and\ \citenamefont
  {Marks}}]{Christie2015}%
  \BibitemOpen
  \bibfield  {author} {\bibinfo {author} {\bibfnamefont {H.}~\bibnamefont
  {Christie}}, \bibinfo {author} {\bibfnamefont {M.}~\bibnamefont {Robinson}},
  \bibinfo {author} {\bibfnamefont {D.}~\bibnamefont {Roach}}, \bibinfo
  {author} {\bibfnamefont {D.}~\bibnamefont {Ross}}, \bibinfo {author}
  {\bibfnamefont {I.}~\bibnamefont {Suarez-Martinez}}, \ and\ \bibinfo {author}
  {\bibfnamefont {N.}~\bibnamefont {Marks}},\ }\href {\doibase
  10.1016/j.carbon.2014.09.031} {\bibfield  {journal} {\bibinfo  {journal}
  {Carbon}\ }\textbf {\bibinfo {volume} {81}},\ \bibinfo {pages} {105–114}
  (\bibinfo {year} {2015})}\BibitemShut {NoStop}%
\bibitem [{\citenamefont {Lumpkin}\ \emph {et~al.}(2008)\citenamefont
  {Lumpkin}, \citenamefont {Smith}, \citenamefont {Blackford}, \citenamefont
  {Thomas}, \citenamefont {Whittle}, \citenamefont {Marks},\ and\ \citenamefont
  {Zaluzec}}]{Lumpkin2008}%
  \BibitemOpen
  \bibfield  {author} {\bibinfo {author} {\bibfnamefont {G.~R.}\ \bibnamefont
  {Lumpkin}}, \bibinfo {author} {\bibfnamefont {K.~L.}\ \bibnamefont {Smith}},
  \bibinfo {author} {\bibfnamefont {M.~G.}\ \bibnamefont {Blackford}}, \bibinfo
  {author} {\bibfnamefont {B.~S.}\ \bibnamefont {Thomas}}, \bibinfo {author}
  {\bibfnamefont {K.~R.}\ \bibnamefont {Whittle}}, \bibinfo {author}
  {\bibfnamefont {N.~A.}\ \bibnamefont {Marks}}, \ and\ \bibinfo {author}
  {\bibfnamefont {N.~J.}\ \bibnamefont {Zaluzec}},\ }\href {\doibase
  10.1103/PhysRevB.77.214201} {\bibfield  {journal} {\bibinfo  {journal} {Phys.
  Rev. B}\ }\textbf {\bibinfo {volume} {77}},\ \bibinfo {pages} {214201}
  (\bibinfo {year} {2008})}\BibitemShut {NoStop}%
\bibitem [{\citenamefont {Trachenko}\ \emph {et~al.}(2006)\citenamefont
  {Trachenko}, \citenamefont {Dove}, \citenamefont {Artacho}, \citenamefont
  {Todorov},\ and\ \citenamefont {Smith}}]{Trachenko2006}%
  \BibitemOpen
  \bibfield  {author} {\bibinfo {author} {\bibfnamefont {K.}~\bibnamefont
  {Trachenko}}, \bibinfo {author} {\bibfnamefont {M.~T.}\ \bibnamefont {Dove}},
  \bibinfo {author} {\bibfnamefont {E.}~\bibnamefont {Artacho}}, \bibinfo
  {author} {\bibfnamefont {I.~T.}\ \bibnamefont {Todorov}}, \ and\ \bibinfo
  {author} {\bibfnamefont {W.}~\bibnamefont {Smith}},\ }\href {\doibase
  http://dx.doi.org/10.1103/PhysRevB.73.174207} {\bibfield  {journal} {\bibinfo
   {journal} {Phys. Rev. B}\ }\textbf {\bibinfo {volume} {73}},\ \bibinfo
  {pages} {174207} (\bibinfo {year} {2006})}\BibitemShut {NoStop}%
\bibitem [{\citenamefont {Xu}\ \emph {et~al.}(2012)\citenamefont {Xu},
  \citenamefont {Buldyrev}, \citenamefont {Stanley},\ and\ \citenamefont
  {Franzese}}]{Xu2012}%
  \BibitemOpen
  \bibfield  {author} {\bibinfo {author} {\bibfnamefont {L.}~\bibnamefont
  {Xu}}, \bibinfo {author} {\bibfnamefont {S.~V.}\ \bibnamefont {Buldyrev}},
  \bibinfo {author} {\bibfnamefont {H.~E.}\ \bibnamefont {Stanley}}, \ and\
  \bibinfo {author} {\bibfnamefont {G.}~\bibnamefont {Franzese}},\ }\href
  {\doibase 10.1103/PhysRevLett.109.095702} {\bibfield  {journal} {\bibinfo
  {journal} {Phys. Rev. Lett.}\ }\textbf {\bibinfo {volume} {109}},\ \bibinfo
  {pages} {095702} (\bibinfo {year} {2012})}\BibitemShut {NoStop}%
\bibitem [{\citenamefont {Elfimova}\ \emph {et~al.}(2013)\citenamefont
  {Elfimova}, \citenamefont {Ivanov},\ and\ \citenamefont
  {Camp}}]{Elfimova2013}%
  \BibitemOpen
  \bibfield  {author} {\bibinfo {author} {\bibfnamefont {E.~A.}\ \bibnamefont
  {Elfimova}}, \bibinfo {author} {\bibfnamefont {A.~O.}\ \bibnamefont
  {Ivanov}}, \ and\ \bibinfo {author} {\bibfnamefont {P.~J.}\ \bibnamefont
  {Camp}},\ }\href {\doibase 10.1103/PhysRevE.88.042310} {\bibfield  {journal}
  {\bibinfo  {journal} {Phys. Rev. E}\ }\textbf {\bibinfo {volume} {88}},\
  \bibinfo {pages} {042310} (\bibinfo {year} {2013})}\BibitemShut {NoStop}%
\bibitem [{\citenamefont {Toxvaerd}(1998)}]{Toxvaerd1998}%
  \BibitemOpen
  \bibfield  {author} {\bibinfo {author} {\bibfnamefont {S.}~\bibnamefont
  {Toxvaerd}},\ }\href {\doibase 10.1103/PhysRevE.58.704} {\bibfield  {journal}
  {\bibinfo  {journal} {Phys. Rev. E}\ }\textbf {\bibinfo {volume} {58}},\
  \bibinfo {pages} {704} (\bibinfo {year} {1998})}\BibitemShut {NoStop}%
\bibitem [{\citenamefont {van~der Spoel}\ \emph {et~al.}(1998)\citenamefont
  {van~der Spoel}, \citenamefont {van Maaren},\ and\ \citenamefont
  {Berendsen}}]{Spoel1998}%
  \BibitemOpen
  \bibfield  {author} {\bibinfo {author} {\bibfnamefont {D.}~\bibnamefont
  {van~der Spoel}}, \bibinfo {author} {\bibfnamefont {P.~J.}\ \bibnamefont {van
  Maaren}}, \ and\ \bibinfo {author} {\bibfnamefont {H.~J.~C.}\ \bibnamefont
  {Berendsen}},\ }\href@noop {} {\bibfield  {journal} {\bibinfo  {journal} {J.
  Chem. Phys.}\ }\textbf {\bibinfo {volume} {108}},\ \bibinfo {pages} {10220}
  (\bibinfo {year} {1998})}\BibitemShut {NoStop}%
\bibitem [{\citenamefont {Robinson}\ \emph {et~al.}(1996)\citenamefont
  {Robinson}, \citenamefont {Zhu}, \citenamefont {Singh},\ and\ \citenamefont
  {Evans}}]{Robinson1996}%
  \BibitemOpen
  \bibfield  {author} {\bibinfo {author} {\bibfnamefont {G.~W.}\ \bibnamefont
  {Robinson}}, \bibinfo {author} {\bibfnamefont {S.~B.}\ \bibnamefont {Zhu}},
  \bibinfo {author} {\bibfnamefont {S.}~\bibnamefont {Singh}}, \ and\ \bibinfo
  {author} {\bibfnamefont {M.~W.}\ \bibnamefont {Evans}},\ }\href@noop {}
  {\emph {\bibinfo {title} {Water in Biology, Chemistry and Physics:
  Experimental Overviews and Computational Methodologies}}},\ \bibinfo {series}
  {World Scientific Series in Contemporary Chemical Physics}, Vol.~\bibinfo
  {volume} {9}\ (\bibinfo  {publisher} {World Scientific},\ \bibinfo {address}
  {Singapore},\ \bibinfo {year} {1996})\BibitemShut {NoStop}%
\bibitem [{\citenamefont {Allen}\ and\ \citenamefont
  {Tildesley}(1987)}]{Allen1987}%
  \BibitemOpen
  \bibfield  {author} {\bibinfo {author} {\bibfnamefont {M.~P.}\ \bibnamefont
  {Allen}}\ and\ \bibinfo {author} {\bibfnamefont {D.~J.}\ \bibnamefont
  {Tildesley}},\ }\href@noop {} {\emph {\bibinfo {title} {Computer Simulation
  of Liquids}}}\ (\bibinfo  {publisher} {Oxford University Press Inc.},\
  \bibinfo {year} {1987})\BibitemShut {NoStop}%
\bibitem [{\citenamefont {Draganic}(2005)}]{Draganic2005}%
  \BibitemOpen
  \bibfield  {author} {\bibinfo {author} {\bibfnamefont {I.}~\bibnamefont
  {Draganic}},\ }\href@noop {} {\bibfield  {journal} {\bibinfo  {journal} {Rad.
  Phys. Chem.}\ }\textbf {\bibinfo {volume} {72}},\ \bibinfo {pages} {181}
  (\bibinfo {year} {2005})}\BibitemShut {NoStop}%
\bibitem [{\citenamefont {Burns}(1989)}]{Burns1989}%
  \BibitemOpen
  \bibfield  {author} {\bibinfo {author} {\bibfnamefont {W.~G.}\ \bibnamefont
  {Burns}},\ }\href {\doibase 10.1038/339515c0} {\bibfield  {journal} {\bibinfo
   {journal} {Nature}\ }\textbf {\bibinfo {volume} {339}},\ \bibinfo {pages}
  {515} (\bibinfo {year} {1989})}\BibitemShut {NoStop}%
\bibitem [{\citenamefont {Sims}(2006)}]{Sims2006}%
  \BibitemOpen
  \bibfield  {author} {\bibinfo {author} {\bibfnamefont {H.}~\bibnamefont
  {Sims}},\ }\href {\doibase 10.1016/j.radphyschem.2006.01.010} {\bibfield
  {journal} {\bibinfo  {journal} {Rad. Phys. Chem.}\ }\textbf {\bibinfo
  {volume} {75}},\ \bibinfo {pages} {1047} (\bibinfo {year}
  {2006})}\BibitemShut {NoStop}%
\bibitem [{\citenamefont {Kolesnichenko}(1975)}]{Kolesnichenko1975}%
  \BibitemOpen
  \bibfield  {author} {\bibinfo {author} {\bibfnamefont {Y.}~\bibnamefont
  {Kolesnichenko}},\ }\href@noop {} {\bibfield  {journal} {\bibinfo  {journal}
  {Nuclear Fusion}\ }\textbf {\bibinfo {volume} {15}},\ \bibinfo {pages} {35}
  (\bibinfo {year} {1975})}\BibitemShut {NoStop}%
\bibitem [{\citenamefont {Martino}\ \emph {et~al.}(2006)\citenamefont
  {Martino}, \citenamefont {de~la Mora},\ and\ \citenamefont
  {Yoshida}}]{Martino2006}%
  \BibitemOpen
  \bibfield  {author} {\bibinfo {author} {\bibfnamefont {W.}~\bibnamefont
  {Martino}}, \bibinfo {author} {\bibfnamefont {F.}~\bibnamefont {de~la Mora}},
  \ and\ \bibinfo {author} {\bibfnamefont {Y.}~\bibnamefont {Yoshida}},\
  }\href@noop {} {\bibfield  {journal} {\bibinfo  {journal} {Green Chemistry}\
  }\textbf {\bibinfo {volume} {8}},\ \bibinfo {pages} {390} (\bibinfo {year}
  {2006})}\BibitemShut {NoStop}%
\bibitem [{\citenamefont {Zenkievicz}(2007)}]{Zenkievicz2007}%
  \BibitemOpen
  \bibfield  {author} {\bibinfo {author} {\bibfnamefont {M.}~\bibnamefont
  {Zenkievicz}},\ }\href@noop {} {\bibfield  {journal} {\bibinfo  {journal} {J.
  Achiev. Mat. Manufact. Eng.}\ }\textbf {\bibinfo {volume} {2}},\ \bibinfo
  {pages} {43} (\bibinfo {year} {2007})}\BibitemShut {NoStop}%
\bibitem [{\citenamefont {Byung}\ and\ \citenamefont {Jung}(2008)}]{Byung2008}%
  \BibitemOpen
  \bibfield  {author} {\bibinfo {author} {\bibfnamefont {M.}~\bibnamefont
  {Byung}}\ and\ \bibinfo {author} {\bibfnamefont {H.}~\bibnamefont {Jung}},\
  }\href@noop {} {\bibfield  {journal} {\bibinfo  {journal} {Appl. Phys.
  Lett.}\ }\textbf {\bibinfo {volume} {93}},\ \bibinfo {pages} {244105}
  (\bibinfo {year} {2008})}\BibitemShut {NoStop}%
\bibitem [{\citenamefont {Weon}\ \emph {et~al.}(2008)\citenamefont {Weon},
  \citenamefont {Je}, \citenamefont {Hwu},\ and\ \citenamefont
  {Margaritondo}}]{Weon2008}%
  \BibitemOpen
  \bibfield  {author} {\bibinfo {author} {\bibfnamefont {B.~M.}\ \bibnamefont
  {Weon}}, \bibinfo {author} {\bibfnamefont {J.~H.}\ \bibnamefont {Je}},
  \bibinfo {author} {\bibfnamefont {Y.}~\bibnamefont {Hwu}}, \ and\ \bibinfo
  {author} {\bibfnamefont {G.}~\bibnamefont {Margaritondo}},\ }\href@noop {}
  {\bibfield  {journal} {\bibinfo  {journal} {Phys. Rev. Lett.}\ }\textbf
  {\bibinfo {volume} {100}},\ \bibinfo {pages} {217403} (\bibinfo {year}
  {2008})}\BibitemShut {NoStop}%
\bibitem [{\citenamefont {Bogolyubov}(1962)}]{Bogolyubov1962}%
  \BibitemOpen
  \bibfield  {author} {\bibinfo {author} {\bibfnamefont {N.}~\bibnamefont
  {Bogolyubov}},\ }\enquote {\bibinfo {title} {Studies in statistical
  mechanics},}\ \ (\bibinfo  {publisher} {North-Holland},\ \bibinfo {year}
  {1962})\ Chap.\ \bibinfo {chapter} {Problems of dynamical theory in
  statistical physics}, p.~\bibinfo {pages} {5}\BibitemShut {NoStop}%
\bibitem [{\citenamefont {Bulavin}\ \emph {et~al.}(2016)\citenamefont
  {Bulavin}, \citenamefont {Cherevko}, \citenamefont {Gavryushenko},
  \citenamefont {Sysoev},\ and\ \citenamefont {Vlasenko}}]{Bulavin2016}%
  \BibitemOpen
  \bibfield  {author} {\bibinfo {author} {\bibfnamefont {L.~A.}\ \bibnamefont
  {Bulavin}}, \bibinfo {author} {\bibfnamefont {K.~V.}\ \bibnamefont
  {Cherevko}}, \bibinfo {author} {\bibfnamefont {D.~A.}\ \bibnamefont
  {Gavryushenko}}, \bibinfo {author} {\bibfnamefont {V.~M.}\ \bibnamefont
  {Sysoev}}, \ and\ \bibinfo {author} {\bibfnamefont {T.~S.}\ \bibnamefont
  {Vlasenko}},\ }\href@noop {} {\bibfield  {journal} {\bibinfo  {journal}
  {Phys. Rev. E}\ }\textbf {\bibinfo {volume} {93}},\ \bibinfo {pages} {032133}
  (\bibinfo {year} {2016})}\BibitemShut {NoStop}%
\bibitem [{\citenamefont {Temperley}\ \emph {et~al.}(1968)\citenamefont
  {Temperley}, \citenamefont {Rowlinson},\ and\ \citenamefont
  {Rushbrooke}}]{Temperley1968}%
  \BibitemOpen
  \bibinfo {editor} {\bibfnamefont {H.~N.~V.}\ \bibnamefont {Temperley}},
  \bibinfo {editor} {\bibfnamefont {J.~S.}\ \bibnamefont {Rowlinson}}, \ and\
  \bibinfo {editor} {\bibfnamefont {G.~S.}\ \bibnamefont {Rushbrooke}},\ eds.,\
  \href@noop {} {\emph {\bibinfo {title} {Physics of Simple Liquids}}}\
  (\bibinfo  {publisher} {North-Holland Publishing Company},\ \bibinfo
  {address} {Amsterdam},\ \bibinfo {year} {1968})\BibitemShut {NoStop}%
\bibitem [{\citenamefont {Berendsen}\ \emph {et~al.}(1981)\citenamefont
  {Berendsen}, \citenamefont {Postma}, \citenamefont {van Gunsteren},\ and\
  \citenamefont {Hermans}}]{Berendsen1981}%
  \BibitemOpen
  \bibfield  {author} {\bibinfo {author} {\bibfnamefont {H.}~\bibnamefont
  {Berendsen}}, \bibinfo {author} {\bibfnamefont {J.}~\bibnamefont {Postma}},
  \bibinfo {author} {\bibfnamefont {W.}~\bibnamefont {van Gunsteren}}, \ and\
  \bibinfo {author} {\bibfnamefont {J.}~\bibnamefont {Hermans}},\ }\enquote
  {\bibinfo {title} {Interaction models for water in relation to protein
  hydration},}\ \ (\bibinfo  {publisher} {Reidel},\ \bibinfo {address}
  {Dordrecht},\ \bibinfo {year} {1981})\ pp.\ \bibinfo {pages}
  {331--342}\BibitemShut {NoStop}%
\bibitem [{\citenamefont {Wu}\ \emph {et~al.}(2006)\citenamefont {Wu},
  \citenamefont {Tepper},\ and\ \citenamefont {Voth}}]{Wu2006}%
  \BibitemOpen
  \bibfield  {author} {\bibinfo {author} {\bibfnamefont {Y.}~\bibnamefont
  {Wu}}, \bibinfo {author} {\bibfnamefont {H.~L.}\ \bibnamefont {Tepper}}, \
  and\ \bibinfo {author} {\bibfnamefont {G.~A.}\ \bibnamefont {Voth}},\ }\href
  {\doibase 10.1063/1.2136877} {\bibfield  {journal} {\bibinfo  {journal} {J.
  Chem. Phys.}\ }\textbf {\bibinfo {volume} {124}},\ \bibinfo {pages} {024503}
  (\bibinfo {year} {2006})}\BibitemShut {NoStop}%
\bibitem [{\citenamefont {Todorov}\ \emph {et~al.}(2006)\citenamefont
  {Todorov}, \citenamefont {Smith}, \citenamefont {Trachenko},\ and\
  \citenamefont {Dove}}]{Todorov2006}%
  \BibitemOpen
  \bibfield  {author} {\bibinfo {author} {\bibfnamefont {I.}~\bibnamefont
  {Todorov}}, \bibinfo {author} {\bibfnamefont {W.}~\bibnamefont {Smith}},
  \bibinfo {author} {\bibfnamefont {K.}~\bibnamefont {Trachenko}}, \ and\
  \bibinfo {author} {\bibfnamefont {M.}~\bibnamefont {Dove}},\ }\href {\doibase
  10.1039/B517931A} {\bibfield  {journal} {\bibinfo  {journal} {J. Mater.
  Chem.}\ }\textbf {\bibinfo {volume} {16}},\ \bibinfo {pages} {1911} (\bibinfo
  {year} {2006})}\BibitemShut {NoStop}%
\bibitem [{\citenamefont {Lorentz}(1881)}]{Lorentz1881}%
  \BibitemOpen
  \bibfield  {author} {\bibinfo {author} {\bibfnamefont {H.~A.}\ \bibnamefont
  {Lorentz}},\ }\href {\doibase 10.1002/andp.18812480110} {\bibfield  {journal}
  {\bibinfo  {journal} {Ann. Phys.}\ }\textbf {\bibinfo {volume} {248}},\
  \bibinfo {pages} {127–136} (\bibinfo {year} {1881})}\BibitemShut {NoStop}%
\bibitem [{\citenamefont {Berthelot}(1898)}]{Berthelot1898}%
  \BibitemOpen
  \bibfield  {author} {\bibinfo {author} {\bibfnamefont {D.}~\bibnamefont
  {Berthelot}},\ }\href@noop {} {\bibfield  {journal} {\bibinfo  {journal} {C.
  R. Acad. Sci.}\ }\textbf {\bibinfo {volume} {126}},\ \bibinfo {pages} {1703–
  1855} (\bibinfo {year} {1898})}\BibitemShut {NoStop}%
\bibitem [{\citenamefont {Zielkiewicz}(2005)}]{Zielkiewicz2005}%
  \BibitemOpen
  \bibfield  {author} {\bibinfo {author} {\bibfnamefont {J.}~\bibnamefont
  {Zielkiewicz}},\ }\href {\doibase http://dx.doi.org/10.1063/1.2018637}
  {\bibfield  {journal} {\bibinfo  {journal} {J. Chem. Phys.}\ }\textbf
  {\bibinfo {volume} {123}},\ \bibinfo {pages} {104501} (\bibinfo {year}
  {2005})}\BibitemShut {NoStop}%
\bibitem [{\citenamefont {Lyubartsev}\ and\ \citenamefont
  {Laaksonen}(1996)}]{Lyubartsev1996}%
  \BibitemOpen
  \bibfield  {author} {\bibinfo {author} {\bibfnamefont {A.~P.}\ \bibnamefont
  {Lyubartsev}}\ and\ \bibinfo {author} {\bibfnamefont {A.}~\bibnamefont
  {Laaksonen}},\ }\href@noop {} {\bibfield  {journal} {\bibinfo  {journal} {J.
  Phys. Chem.}\ }\textbf {\bibinfo {volume} {100}},\ \bibinfo {pages} {16410}
  (\bibinfo {year} {1996})}\BibitemShut {NoStop}%
\bibitem [{\citenamefont {Tang}\ \emph {et~al.}(2006)\citenamefont {Tang}, ,\
  and\ \citenamefont {Ford}}]{Tang2006}%
  \BibitemOpen
  \bibfield  {author} {\bibinfo {author} {\bibfnamefont {H.~Y.}\ \bibnamefont
  {Tang}}, , \ and\ \bibinfo {author} {\bibfnamefont {I.~J.}\ \bibnamefont
  {Ford}},\ }\href {\doibase 10.1063/1.2357147} {\bibfield  {journal} {\bibinfo
   {journal} {J. Chem. Phys.}\ }\textbf {\bibinfo {volume} {125}},\ \bibinfo
  {pages} {144316} (\bibinfo {year} {2006})}\BibitemShut {NoStop}%
\bibitem [{\citenamefont {Lemmon}\ \emph {et~al.}(2007)\citenamefont {Lemmon},
  \citenamefont {Huber},\ and\ \citenamefont {McLinden}}]{Lemmon2007}%
  \BibitemOpen
  \bibfield  {author} {\bibinfo {author} {\bibfnamefont {E.}~\bibnamefont
  {Lemmon}}, \bibinfo {author} {\bibfnamefont {M.}~\bibnamefont {Huber}}, \
  and\ \bibinfo {author} {\bibfnamefont {M.}~\bibnamefont {McLinden}},\
  }\href@noop {} {\emph {\bibinfo {title} {REFPROP: Reference fluid
  thermodynamic and transport properties. Version 8.0}}},\ NIST standard
  reference database\ (\bibinfo  {publisher} {NIST},\ \bibinfo {address}
  {Gaithersburg},\ \bibinfo {year} {2007})\BibitemShut {NoStop}%
\bibitem [{\citenamefont {Guillot}(2002)}]{Guillot2002}%
  \BibitemOpen
  \bibfield  {author} {\bibinfo {author} {\bibfnamefont {B.}~\bibnamefont
  {Guillot}},\ }\href {\doibase 10.1016/S0167-7322(02)00094-6} {\bibfield
  {journal} {\bibinfo  {journal} {J. Mol. Liq.}\ }\textbf {\bibinfo {volume}
  {101}},\ \bibinfo {pages} {219} (\bibinfo {year} {2002})}\BibitemShut
  {NoStop}%
\bibitem [{\citenamefont {Guevara-Carrion}\ \emph {et~al.}(2011)\citenamefont
  {Guevara-Carrion}, \citenamefont {Vrabec},\ and\ \citenamefont
  {Hasse}}]{GuevaraCarrion2011}%
  \BibitemOpen
  \bibfield  {author} {\bibinfo {author} {\bibfnamefont {G.}~\bibnamefont
  {Guevara-Carrion}}, \bibinfo {author} {\bibfnamefont {J.}~\bibnamefont
  {Vrabec}}, \ and\ \bibinfo {author} {\bibfnamefont {H.}~\bibnamefont
  {Hasse}},\ }\href {\doibase 10.1063/1.3515262} {\bibfield  {journal}
  {\bibinfo  {journal} {J. Chem. Phys.}\ }\textbf {\bibinfo {volume} {134}},\
  \bibinfo {pages} {074508} (\bibinfo {year} {2011})}\BibitemShut {NoStop}%
\end{thebibliography}%

\end{document}